\documentclass[aps,nofootinbib,notitlepage,11pt]{revtex4-1}
\usepackage{slashed}
\usepackage{amsmath,amsfonts,amssymb}
\usepackage{graphicx}
\usepackage{bm}
\usepackage[colorlinks,linkcolor=blue,citecolor=blue]{hyperref}
\usepackage[usenames,dvipsnames,svgnames,table]{xcolor}

\newcommand{\di}{\text{d}}
\newcommand{\be}{\begin{equation}}
\newcommand{\ee}{\end{equation}}
\newcommand{\bea}{\begin{eqnarray}}
\newcommand{\eea}{\end{eqnarray}}
\newcommand{\ba}{\begin{eqnarray}}
\newcommand{\ea}{\end{eqnarray}}

\newcommand{\Tr}{\mbox{Tr}\;}

\begin{document}

\title{Polarized $\phi$ meson rates with viscous corrections at RHIC}

\author{Eduardo Grossi}
\email{eduardo.grossi@unifi.it }
\affiliation{Dipartimento di Fisica, Universit\`a di Firenze and INFN Sezione di Firenze, via G. Sansone 1,
50019 Sesto Fiorentino, Italy}

\author{Andrea Palermo}
\email{andrea.palermo@stonybrook.edu}
\affiliation{Center for Nuclear Theory, Department of Physics and Astronomy, Stony Brook University, Stony Brook, New York 11794--3800, USA}

\author{Ismail Zahed}
\email{ismail.zahed@stonybrook.edu}
\affiliation{Center for Nuclear Theory, Department of Physics and Astronomy, Stony Brook University, Stony Brook, New York 11794--3800, USA}

\begin{abstract}
We discuss the emission of $\phi$ mesons with longitudinal and transverse polarization in ultra-relativistic heavy ion collisions.
In the hadronic phase and in leading order in the kaon diluteness, $\phi$ emission is isotropic and driven by
the flavor singlet and octet vector spectral functions.  At next to leading order, the emissivities receive
a non-isotropic correction from the flavor octet spectral function, and the non-equilibrium contributions originating from the shear and bulk viscosities. Implications for the $\phi$ meson alignment in heavy-ion collisions are discussed.
\end{abstract}

\maketitle

\section{Introduction}
Spin polarization is becoming a well-established observable for investigating the quark-gluon plasma generated in heavy ion collisions. Early on, it was proposed that the orbital angular momentum in non-central collisions could be transferred to partons and eventually to the spin of final state particles. Such a process would result in the spin ``vector'' polarization of Dirac fermions, such as the $\Lambda$ hyperon, and the ``tensor'' polarization of spin-one mesons like $\phi$ and $J/\psi$ \cite{Liang:2004ph,Liang:2004xn,Becattini:2007sr}. For a recent review of these phenomena, see Ref.~\cite{Becattini:2024uha}. The tensor polarization of vector mesons measured in strong decays is generally characterized by the \emph{spin alignment}, which quantifies the deviation from an isotropic spin distribution. 

Both spin polarization and vector meson spin alignment are accessible experimentally
\cite{STAR:2017ckg,STAR:2022fan,Kundu:2019bwr,ALICE:2022dyy}.  
A local equilibrium model proves successful in describing the $\Lambda$ polarization~\cite{Palermo:2024tza,Becattini:2021suc,Becattini:2021iol,Palermo:2023cup,Liu:2021uhn,Fu:2021pok,Yi:2021ryh}. According to this model, polarization is generated by the collective motion of the Quark Gluon Plasma (QGP) created in the collisions, the gradients of temperature and velocity being the source of polarization. In the case of alignment, however, the predictions from the thermal model are too small to describe the experimental data \cite{Becattini:2016gvu,Dong:2023cng}.  This observation spurred a great deal of theoretical work \cite{Li:2022vmb,Muller:2021hpe,Yang:2017sdk,Yang:2017sdk,Xia:2020tyd,Wagner:2022gza,Wagner:2023cct,Dong:2023cng,Kumar:2023ojl,Sheng:2024kgg,Yang:2024ejk,Li:2024qae}. The most successful approach so far identifies the source of alignment in local fluctuations of the strong force field, whose magnitude is extracted from the data \cite{Sheng:2019kmk,Sheng:2022wsy,Sheng:2023urn}.  

Here we study the spin alignment of $\phi$ mesons due to the interaction with kaons \cite{Park:2022ayr}. The formalism employed is based on a generalization of \cite{Yamagishi:1995kr} to $SU(3)$. Such a formalism was developed as an alternative to Chiral Effective theory, where the strictures of broken chiral symmetry are enforced in the form of general Ward identities using a chiral reduction scheme, by generalizing the famed current algebra method to the physical pion point. In this sense, our approach is similar to a recent paper, where the alignment of $\rho$ mesons interacting with pions was studied in chiral effective theory \cite{Yin:2024dnu}. 
The chiral reduction approach provides the most general framework for the analysis of the electromagnetic emissivities at collider energies~\cite{Steele:1996su,Dusling:2006yv,Dusling:2007su,Kim:2016ylr}, including
the effects of non-equilibrium and viscosities~\cite{Liu:2017fib}. Therefore, it is natural to use it for the analysis of the light vector meson emissivities, e.g. $\rho, \phi$,  at collider energies.

The physical mechanism to explain the spin-alignment for vector meson is different with respect to the polarization of $\Lambda$ particles. 
The spin alignment is given by the sizeable difference of the production 
of the longitudinally polarized vector meson to the transverse one. 
The different production rates of the two channels can be easily explained by the vector meson nature of their interaction. 
In a thermal bath 
the temperature breaks the Lorentz invariance of the rate, introducing a preferred frame to the vector meson produced.
Consequently, the longitudinal and the transverse rates differ naturally due to thermal corrections. 

The paper's outline is as follows, in section~\ref{SEC2} we detail the thermal equilibrium rates for the emission of $\phi$ mesons with different polarizations. 
At leading order, the rates are isotropic as expected from a resonance gas. At next-to-leading order in the kaon density, the $\phi$ emission rates are shown to be anisotropic,
due to the strong interactions with the kaons through pertinent spectral functions. 
In section~\ref{SEC3} we derive the corrections to the emission rates caused by the 
bulk and shear viscosities of the underlying hadronic fluid. In section~\ref{SEC4}, we describe the spectral functions needed for the analysis. After that, the rates are first evaluated for a static fluid in thermal equilibrium, where the various corrections to the isotropic results are analyzed. The rates are then re-evaluated in a Bjorken fluid and compared to the recently reported STAR data. Our conclusions are in section~\ref{SEC5}, and some supporting details are given in the Appendix.

%%%%%%%%%%%%%%%%%%%%END%%%%%%%%%%%%%%%%%%%%%%%

\section{Emission rates in equilibrium}
\label{SEC2}
To analyze the emission of $\phi$ mesons in an interacting thermal hadronic fluid, 
we will use vector dominance to tie the $\phi$ emissivities to the strangeness
current-current correlator. In equilibrium,  
stable pions and kaons asymptotically dominate the thermal hadronic fluid in the baryon-free region, that is in high energy heavy ion collisions. In this regime, we make
use of the chiral reduction scheme developed in~\cite{Yamagishi:1995kr},  as used for the reduction of the thermal electromagnetic emissivities in~\cite{Steele:1996su,Dusling:2006yv,Dusling:2007su,Kim:2016ylr}. In leading order,
the $\phi$ emission rates are those of a resonance gas, with corrections at next-to-leading orders in the density of stable particles, stemming mostly from the strong interaction with kaons. These interactions are captured by pertinent spectral functions under the general strictures of spontaneously broken chiral symmetry. 

\subsection{Rate}
The emission rate of polarized $\phi$ mesons from a hadronic fireball in equilibrium is given by
\begin{equation}\label{eq:number of phis}
N_\phi= \sum_{I,F} |\mathcal{M}|^2 \times \frac{e^{-\beta E_I}}{Z} \times \frac{V d^3 q }{ (2\pi)^3}, \quad \text{with} \quad q^2=m_\phi^2 ,
\end{equation}
where $Z$ is the partition function at a finite temperature of the gas in the initial state, $V$ is the volume of the fireball, $q$ is the momenta of the emitted $\phi$ particle, and the sum is extended on the initial and final state of the corresponding process. 
The square amplitude $|\mathcal{M}|^2$ can be obtained from 
\begin{align}
 |\mathcal{M}|^2\equiv | \langle F \phi |\mathcal{S}| I \rangle |^2 ,
\end{align}
and represents the process of the emission of a $\phi$ and a generic final state $F$ form an initial state $I$.
The relevant $\mathcal{S}$-matrix element for this process can be approximated as
\begin{equation}
\mathcal{S} = \int \di^4 x \,G_V \phi_\mu \bar{s} \gamma^{\mu} s \equiv \int \di^4 x G_V \phi_\mu J_{s}^{\mu},
\end{equation}
where $\phi_\mu$ is the $\phi$ field, $J^{\mu}_s$ is the current of $\bar s s $ quark, and $G_V$ is the effective coupling. 
The matrix element can be written as 
\begin{equation}\label{eq:matrix elem}
|\mathcal{M}|^2 = | \langle F \phi |\int d^4 x\;  G_V\phi_\mu J^{\mu}_s  |  I \rangle |^2 .
\end{equation}
The $\phi$ state can be defined as  
\begin{equation}
 \langle \phi(q,\sigma)  |   J^{\mu}_s |0\rangle =\frac{ f_\phi m_\phi \epsilon_\sigma^{\mu}(q)  }{\sqrt{2 E^\phi_qV} },
\end{equation}
where $m_\phi$ is the on-shell mass of the $\phi$ meson, the energy is $E_q^\phi=\sqrt{\vec q^2+m_\phi^2}$, $f_\phi$ is the $\phi$'s decay constant and $\epsilon_\sigma^\mu$ is the polarization vector associated with the spin state $\sigma$ of the meson. 
As a vector meson, the spin can have only three independent components conventionally parametrized by the quantum number $\sigma=-1,0,1$.  
The basis of polarization vectors obeys
\begin{equation}
    q_\mu\epsilon_\sigma^\mu(q) =0, \quad \sum_{\sigma=1}^3\epsilon^\mu_\sigma(q){\epsilon^*}^\nu_\sigma(q) = \frac{q^{\mu}q^{\nu}}{m^2_\phi} - g^{\mu\nu}, \quad \epsilon_\sigma^\mu(q) {\epsilon^*_{\sigma'}}_\mu(q) = -\delta_{\sigma,\sigma'}.
\end{equation}
The explicit definition of these vectors is given in appendix \ref{app:integrals} in the rest frame of the $\phi$.
The matrix element in eq. \eqref{eq:matrix elem} can be reduced further
\begin{equation}
\langle F \phi |\int d^4 x\;  G_V \phi_\mu J^{\mu}_s  |  I \rangle= G_V \langle \phi(q,\sigma)  |   J^{\mu}_s |0\rangle \int  d^4 x\; e^{iq\cdot x} \langle F |  \phi_\mu(x)   |  I \rangle, 
\end{equation}

Using eqs. \eqref{eq:number of phis} and \eqref{eq:matrix elem}, the number of produced $\phi$s with spin states $\sigma$ and $\sigma'$ is therefore given by
\begin{align}
N^\phi_{\sigma,\sigma'}=&G_V^2(f_\phi m_\phi )^2\epsilon^{\mu}_\sigma(q) {\epsilon^*}^{\nu}_{\sigma'}(q) \frac{ d^3 q }{ (2\pi)^32 E^\phi_q  } \nonumber\\
& \times \sum_{I,F} \frac{e^{-\beta E_I}}{Z} 
\int  d^4 y\;  \int  d^4 x\; e^{iq\cdot (x-y)} \langle F |  \phi_\mu(x)   |  I \rangle 
 \langle I |  \phi_\nu(y)   |  F \rangle .
\end{align}
The Boltzmann weight can be written using $E_I=E_F+q^0$ to allow the use of the completeness relation on the initial state $|I\rangle$. Using also transnational invariance, we obtain
\begin{equation}
N^\phi_{\sigma,\sigma'}=G_V^2(f_\phi m_\phi )^2\epsilon^{\mu}_\sigma(q) {\epsilon^*}^{\nu}_{\sigma'}(q) \frac{ d^3 q e^{-\beta q^0} }{ (2\pi)^32 E^\phi_q  } \sum_{F}\frac{e^{-\beta E_F}}{Z} 
V_4\;  \int  d^4 x\; e^{iq\cdot x} \langle F |  \phi_\mu(x)   \phi_\nu(0)   |  F \rangle , 
\end{equation}
where $V_4$ is the total four-volume of the system.
Therefore, the number of particles produced is proportional to the four-volume fireball. It is useful to consider the rate since it is independent of this explicit volume factor. By definition
\begin{equation}
dR_{\sigma,\sigma'} = N^\phi_{\sigma,\sigma'}/ V_4 =G_V^2(f^2_\phi m_\phi )^2\epsilon^{\mu}_\sigma(q) {\epsilon^*}^{\nu}_{\sigma'}(q)  e^{-\beta q^0} 
  \int  d^4 x\; e^{iq\cdot x}\langle   \phi_\mu(x)   \phi_\nu(0)   \rangle  \frac{ d^3 q  }{ (2\pi)^32 E^\phi_q }  , 
\end{equation}
where we have introduced the notation $\langle O \rangle$ to describe the thermal average carried over the (final) thermal states.

The correlation function of the mesonic field can be approximated using the vector meson dominance model \cite{Sakurai:1960ju}. 
Accordingly, the quark  current is proportional to the vector meson field 
\begin{equation}
  J^{s}_{\mu} = \frac{m_\phi^2}{G_V}\phi_\mu + \cdots   
\end{equation}
where we neglect higher-order resonances.
Therefore, the vector meson dominance can be used to write the correlation function as
\begin{equation}
 \langle \phi_{\mu}(x)\phi_\nu(0)\rangle= \left(\frac{G_V}{m_\phi^2}\right)^2 \langle J^s_\mu(x) J^s_{\nu}(0)\rangle . 
\end{equation}
Within this approximation, the Lorentz-invariant differential rate reads
\begin{align}
2E_q\frac{\di R_{\sigma,\sigma'}}{\di^3q} &= \frac{G_V^4f^2_\phi}{(2\pi)^3 m_\phi^2}\epsilon^{\mu}_\sigma(q) {\epsilon^*}^{\nu}_{\sigma'}(q)  e^{-\beta q^0} 
  \int  d^4 x\; e^{iq\cdot x}\langle   J^s_\mu(x) J^s_{\nu}(0)  \rangle  \nonumber\\
  &= 
  \frac{G_V^4f^2_\phi}{(2\pi)^3 m_\phi^2}\epsilon^{\mu}_\sigma(q) {\epsilon^*}^{\nu}_{\sigma'}(q)  e^{-\beta q^0} 
  W_{\mu\nu}^{>}(q), 
\end{align}
where in the last equation, we have used the definition of the Wightman function $W^{>}_{\mu\nu}$
\begin{equation}
    W^{>}_{\mu\nu}(q)= \int  d^4 x\; e^{iq\cdot x}\langle   J_\mu(x) J_{\nu}(0)  \rangle 
\end{equation}
Instead of using the Wightman function,  it is more convenient to use the Feynman propagator.
The latter is just the time-ordered two-point function, defined as 
\begin{equation}
W^F_{\mu\nu}(q)=\int d^4 x \langle T  J_\mu(x) J_{\nu}(0)  \rangle=\int d^4 x \theta(x^0) \langle   J_\mu(x) J_{\nu}(0)  \rangle + \theta(-x^0) \langle  J_{\nu}(0)J_\mu(x)  \rangle.
\end{equation}
Its relation to the Wightman is obtained using the KMS condition and reads:
\begin{equation}
    W_{\mu\nu}^>(q)e^{-q_0\beta}= \frac{2\text{Im} \left(iW^F_{\mu\nu}\right)}{e^{\beta q^0}+1}.
\end{equation}
In terms of the Feynman propagator, the invariant production rate for unit of phase space is given by
\begin{equation}\label{eq:spin dep rate}
2E_q\frac{\di R_{\sigma,\sigma'}}{\di^3q} = \frac{1}{e^{\beta q^0}+1}\frac{2G_V^4f^2_\phi}{(2\pi)^3m_\phi^2}\epsilon^{\mu}_{\sigma}(q) {\epsilon^*}^{\nu}_{\sigma'}(q)   \text{Im}\left(iW^F_{\mu\nu}(q) \right).
\end{equation}
The total, spin-independent rate can now be computed summing over $\sigma$ and $\sigma'$. Integrals of the rate define momentum-dependent multiplicities, $N_{\sigma,\sigma'}(q)$. From those formulae, we can also compute the spin density matrix of the $\phi$, which is defined as:
\begin{equation}
    \rho_{\sigma\sigma'}(q)=\frac{N_{\sigma,\sigma'}(q)}{\sum_{\sigma\sigma'}N_{\sigma,\sigma'}(q)}=\frac{N_{\sigma,\sigma'}(q)}{N_{tot}(q)}.
\end{equation}
The alignment parameter studied in experiments represents the deviation of $\rho_{00}$ from it's isotropic value $1/3$, where $\rho_{00}$ refers to the fraction of longitudinally polarized mesons.

We will now compute the Feynman propagator considering a resonance gas and corrections due to interactions and dissipation. These corrections will come to first order in the kaon diluteness, as we detail below.

\subsection{Virial expansion of the Feynman propagator }
The Feynman propagator in a thermal state is:
\begin{equation}
    W^F_{\mu\nu}(x) =\frac{1}{Z} \Tr \left[e^{-\beta H} T(\phi_\mu(x) \phi_\nu(0)) \right]
\end{equation}
In the hadron regime, the trace runs over the strong-force stable asymptotic multi-particle states of the gas, which are mainly composed of pions, kaons, and nucleons. Dealing with strange currents, the leading term contributing to the virial expansion is the 1-kaon state
\begin{align}
    W^F_{\mu\nu}(q)
    =&\int\di^4x e^{iq\cdot x}\langle 0|T ({J_s}_\mu(x){J_s}_\nu(0)) |0\rangle\nonumber\\
    &+\sum_a 
    \int \frac{\di^3k}{(2\pi)^3} \frac{n(k)}{2k^0}\int\di^4x e^{iqx}\langle K^a(k)|T (J_{s\mu}(x)J_{s\nu}(0) )|K^a(k)\rangle_{\text{conn.}}+\dots\\
    =&W_{\mu\nu}^F(q)_0+W_{\mu\nu}^F(q)_1+\dots
\end{align}
where $k$ is the on-shell kaon momentum and $n(k)=[e^{\beta(k^0- \mu)}-1]^{-1}$ is the kaon distribution function, and $\mu$ the  kaon chemical potential.  
The series of the Feynamn propagator is developed in terms of the density of the asymptotic state, in this case the 
kaons density.  We  have introduced the notation $W_{\mu\nu}^F(q)_{0,1}$ for the zeroth and leading orders in this parameter. The matrix element of the current is suppressed by a decay constant factor $f_K$, for each kaon state. Therefore, the effective expansion parameter is the \emph{diluteness}, defined (for kaons) as 
\begin{align}
\label{KAPPAK}
\kappa_K=\frac 1{f_K^2}\int \frac{d^3k}{(2\pi)^3}\frac {n(k)}{2k^0}.
\end{align}
\begin{figure}
    \centering
    \includegraphics{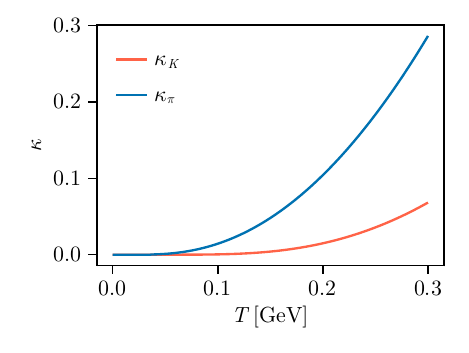}
    \caption{The diluteness parameters of pions $\kappa_\pi$ and kaons $\kappa_K$, as a function of temperature and for vanishing chemical potentials.}
    \label{fig:diluteness}
\end{figure}
In figure \ref{fig:diluteness}, we show the diluteness parameter at vanishing chemical potential as a function of temperature for pions and kaons. We see that the diluteness of kaons and pions is much less than one, making it a good expansion parameter. However, the kaon diluteness is also lower than the pion one due to the larger mass and decay constant, $f_K\approx 1.24 f_\pi$ (with $f_\pi\approx 93$ MeV). This effect makes the corrections to the hadron resonance gas smaller. In the next subsections, we will evaluate the Feynman propagator in the first order of kaon diluteness.

 \subsection{Leading order of $W^F$: resonance gas}
The leading contribution to the Feynman propagator is saturated by the $\phi$ excitations in the vacuum. Therefore the Feynman propagator is related to the $\phi$ spectral function $\Pi_V^\phi (q)$ as
\begin{equation}
 i W_{0F}^{\mu\nu}(q) = i \langle 0 |T(J_s^\mu(q) J_s^{\nu}(0))|0 \rangle_F =(-q^2g^{\mu\nu}+q^\mu q^\nu)\Pi_V^\phi (q).
\end{equation}
Using \eqref{eq:spin dep rate}, together with $\epsilon_\sigma^\mu(q)q_\mu=0$ and $\epsilon_{\sigma}^{\mu}(q) {\epsilon^*_{\sigma'}}_{\mu}(q)=-\delta_{\sigma,\sigma'}$, we readily find:
 \begin{align}
 \label{eq:W0 res gas}
 2E_q\frac{dR^0_{\sigma,\sigma'}}{d^3 q }  
 &=\delta_{\sigma,\sigma'} \bigg[\frac{1}{e^{\beta E^\phi_q}+1}\frac{G_V^4f^2_\phi}{m_\phi^2}   \frac{ 2q^2 }{ (2\pi)^3 }    \text{Im }\Pi_V^\phi(q)\bigg]_{q^2=m_\phi^2},
 \end{align}
 where the rate is to be evaluated for on-shell $\phi$ mesons. The explicit form of the spectral function will be given in a later section, as it is irrelevant to the current discussion. Indeed, we can see that the rate is isotropic in the longitudinal and transverse polarizations.
This is the emission rate expected from an equilibrated hadronic resonance gas. The associated spin density matrix is 
$$\rho^{0}_{\sigma,\sigma'}=\frac{\delta_{\sigma,\sigma'}}{3}.$$

As we now discuss, the isotropy is lifted by hadronic correlations in equilibrium and non-equilibrium viscous effects.

\subsection{Leading corrections of $W^F$ in diluteness}
To evaluate the corrections in the kaon diluteness to the resonance gas result, we will 
use the chiral reduction scheme developed in~\cite{Yamagishi:1995kr,Steele:1996su,Dusling:2006yv,Dusling:2007su,Kim:2016ylr}. For this analysis, we split the strange current into singlet and octet flavor currents
\begin{equation}
\label{JSS}
J_{\mu}^s = \bar s \gamma_{\mu }s =\frac{1}{3} \mathbb{I} -\frac{1}{\sqrt{3} } \lambda_{8}= \frac{1}{\sqrt{6}} \lambda_0 -\frac{1}{\sqrt{3} } \lambda_{8}
\end{equation}
where $\lambda^a$ are a convenient basis of the U(3) Lie algebra (flavor space). 
The square of the current is 
\begin{equation}
J^s J^s = \frac16 J^0 J^0 + \frac13 J^8 J^8+  \frac{1}{3\sqrt{2}} (J^0 J^8+J^8 J^0).
\end{equation}
In what follows, we will ignore the 08-mixing (no leading s-channel resonance). Then we have
\begin{equation}
i W^F_{\mu\nu}=\frac16 \Pi^{00}_{\mu\nu}+\frac13 \Pi^{88}_{\mu\nu}
\end{equation}

With this in mind, the  first correction to the resonance gas is given by
 \bea
 \label{RATEE}
 \frac{dR^1_{L,\perp}}{d^3 q }  =\frac{1}{e^{\beta E^\phi_q}+1}\frac{G_V^4f^2_\phi}{m_\phi^2}\epsilon_{L,\perp}^{\mu}(q) \epsilon_{L,\perp}^{\ast \nu}(q)   \frac{ 1  }{ (2\pi)^3 E^\phi_q }  \int \frac{\di^3k}{(2\pi)^3} \frac{n(k)}{2k^0} \text{Im}\left(i W_{1 F \mu\nu} \right).
 \eea
with 
\bea
i W_{1 F \mu\nu} =&&
(g_{\mu\nu}q^2-q_\mu q_\nu)\,\frac{4}{f_K^2}\,{\rm Im}\Pi_V^{88}(q)
-(g_{\mu\nu}(k\pm q)^2-(k\pm q)_\mu(k\pm q)_\nu)\frac 2{f_K^2}
{\rm Im}\Pi_A^{U}(k\pm q)\nonumber\\
&&+(2k_\mu+q_\mu)(-k_\nu q^2+k\cdot q q_\nu)\frac 4{f_K^2}{\rm Re}\Delta_R(k\pm q)\,{\rm Im}\Pi_V^{88}(q)
\eea 
where the $\pm$ terms are summed over. Here  $\Pi_A^U= \frac 12(\Pi_A^{66}+\Pi_A^{77})$ is the U-spin axial spectral function. For covariantly transverse polarizations $q\cdot\epsilon(q)=0$, we may simplify the contraction
\bea
(2k_\mu+q_\mu)(-k_\nu q^2+k\cdot q q_\nu)\epsilon^\mu(q)\epsilon^{\nu*}(q)=
-2 q^2|k\cdot \epsilon(q)|^2
\eea
which yields
\begin{align}\label{eq:rate1}
\frac{dR^1_{L,\perp}}{d^3 q }  =&\frac{1}{e^{\beta E^\phi_q}+1}\frac{G_V^4f^2_\phi}{m_\phi^2}  \frac{ 1  }{ (2\pi)^3 E^\phi_q } \,
\int \frac{d^3p}{(2\pi)^3}\frac 1{2E^K_p}\frac 1{e^{\beta E^K_p}-1}\nonumber\\
&\times\bigg[\frac {4q^2}{f_K^2}\, {\rm Im}\Pi_V^{88}(q)
-((p\pm q)^2-|\epsilon_{L,\perp}\cdot (p\pm q)|^2)\frac 2{f_K^2}\,  {\rm Im}\Pi_A^{U}((p\pm q))\nonumber\\
&-q^2|\epsilon_{L,\perp}\cdot p|^2\frac{8}{f_K^2}\,{\rm Re}\,\Delta_R(p\pm q)\, {\rm Im}\Pi_V^{88}(q)\bigg]_{q^2=m_\phi^2}
\end{align}
with $E_p^K=\sqrt{\vec p^2+m_K^2}$. The retarded kaon propagator is  given in terms of the PP-values (principal parts)
\begin{align}
{\rm Re}\,\Delta_R(p\pm q)=\frac{\rm PP}{(p+q)^2-m_K^2}+\frac{\rm PP}{(p-q)^2-m_K^2}
\end{align}
In comparison to eq. \eqref{eq:W0 res gas}, the rate in eq. \eqref{eq:rate1} is suppressed by the kaon diluteness factor.

\section{Viscous corrections}
\label{SEC3}
We now analyze the effects of the fluid viscosities on the $\phi$ emission rates when the fluid goes
out of thermal equilibrium. The general framework for addressing such effects was developed for the electromagnetic emissivities in~\cite{Liu:2017fib} and is based on the same methods employed to obtain eq. \eqref{eq:rate1}. 
At the leading order in the kaon diluteness expansion and in the frame where the fluid is at rest, they are given by 
\begin{align}
\label{RATEX}
\frac{dR^{\text{visc}}_{L,\perp}}{d^3 q }  =&\frac{1}{e^{\beta E^\phi_q}+1}\frac{G_V^4f^2_\phi}{m_\phi^2}  \frac{ 1  }{ (2\pi)^3 E^\phi_q } \,  \int \frac{d^3p}{(2\pi)^3}\frac 1{2E^K_p}\frac {e^{\beta E^K_p}}{(e^{\beta E^K_p}-1)^2}\frac {p^2}{E^K_p}
\left(t_\eta\sigma (3{\rm cos}^2\theta_p-1)+\frac{2}{3}t_\zeta\theta\right)\nonumber\\
&\times\bigg[\frac {4q^2}{f_K^2}\, {\rm Im}\Pi_V^{88}(q)
-((p\pm q)^2-|\epsilon_{L,\perp}\cdot (p\pm q)|^2)\frac 2{f_K^2}\,  {\rm Im}\Pi_A^{U}((p\pm q))\nonumber\\
&-q^2|\epsilon_{L,\perp}\cdot p|^2\frac{8}{f_K^2}\,{\rm Re}\,\Delta_R(p\pm q)\, {\rm Im}\Pi_V^{88}(q)\bigg]_{q^2=m_\phi^2}
 \end{align}
The shear $t_\eta$ and bulk $t_\zeta$ time scales are
\begin{align}
t_\eta\approx \frac{\eta}{e+p}=\frac{\eta/s}{T},\qquad \qquad t_\zeta\approx \frac{\zeta}{e+p}=\frac{\zeta/s}{T},\end{align}
where $\eta$ and $\zeta$ are the shear and bulk viscosity, respectively.
The expansion scalar is $\theta= \partial_m\beta_m$, and the shear parameter  $\sigma=\sigma_{ij}\hat k_i\hat k_j$, with $\hat k_i =k_i/|k|$, and the shear tensor $\sigma_{ij}$ is, 
\begin{align}
\sigma_{ij}=\frac 12\bigg(\partial_i\beta_j+\partial_j\beta_i-\frac 23 \delta_{ij}\partial_m\beta_m\bigg)
\end{align}
The full emissivity rates of longitudinal and transverse $\phi$ mesons from an off-equilibrated hadronic gas is
\begin{align}\label{eq:rate tot}
\frac{dR_{L,\perp}}{d^3 q } = \frac{dR^{0}_{L,\perp}}{d^3 q } +\frac{dR^{1}_{L,\perp}}{d^3 q } +\frac{dR^{\text{visc}}_{L,\perp}}{d^3 q } 
+{\cal O}(\kappa_K^2).
\end{align}
As we have shown, the first contribution is that of an equilibrated resonance gas,
the second contribution arises at the next order from the interactions in the resonance gas, and the last contribution
is due to local deviations from equilibrium as captured by the shear and bulk viscosities. All hadronic interactions are described
by spectral functions in the vacuum thanks to the chiral reduction scheme. We emphasize that the latter enforces all the chiral Ward identities
on the pion and kaon mass shell, and bypasses  all the uncertainties related to chiral effective theories, through the use of spectral functions. We can now evaluate numerically the ensuing rates and alignment parameters.

\section{Numerical estimates}
\label{SEC4}
The numerical estimates will be carried out for a static fluid in thermal equilibrium as a first test. We then consider a Bjorken flow~\cite{Bjorken:1982qr}, for a semi-realistic comparison to the recently reported anisotropies by the STAR collaboration at RHIC. We point out, however, that the formulae derived here only hold in a baryon-free fluid, which is expected only in high-energy heavy ion collisions. Baryonic states should be added in the virial expansion to compare with lower energy collisions, which may be done in the future. 

To compute the integrals associated with the interactions and dissipation efficiently, it is convenient to Lorentz transform the integral variable, the kaon momentum, to the rest frame of the $\phi$. This procedure is explained in more detail in appendix \ref{app:integrals}. We now start by specifying the spectral functions to be used.

\subsection{Spectral functions}
\begin{figure}
    \centering
    \includegraphics[scale=1]
    {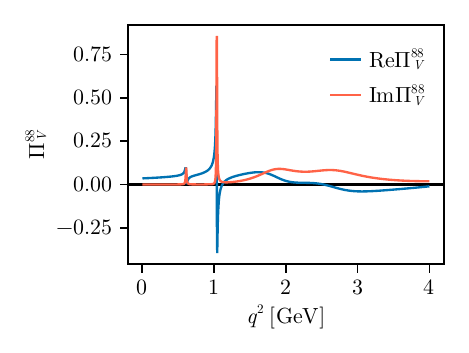}
    \includegraphics[scale=1]
    {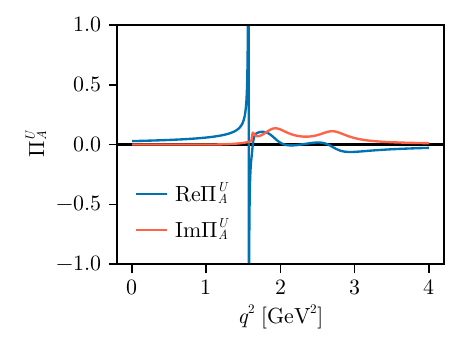}
    \caption{Left panel:
    The vector spectral function $\Pi^{88}_V$ as function of the four momenta $q$
    . Right panel: 
    The axial spectral function $\Pi^{U}_A$ as function of the four momenta $q$
    . }
    \label{fig:spectralfunction}
\end{figure}

For a particle with mass $m$, the vector and axial SU(2) iso-spectral functions are written as:
\begin{align}
\label{PIVA}
\Pi_V^I(q^2)=\frac {f_V^2}{q^2}({\bf F}_V(q^2)-1), \qquad 
\Pi_A^I(q^2)=\frac{f_V^2}{m_V^2-q^2-im_V\Gamma(q^2)},
\end{align}
The empirical vector form factor ${\bf F}_V$ and decay widths $\Gamma$ are parametrized as follows:
\begin{align}
{\bf F}_V(q^2)=&\frac{m_V^2}{m_V^2-q^2-im_V \Gamma(q^2)}\nonumber\\
\Gamma(q^2)=&\theta(q^2-M^2_{BR})\Gamma_{0}\frac{m_V}{\sqrt{q^2}}\bigg(\frac{q^2-M^2_{BR}}{m_V^2-M^2_{BR}}\bigg)^{\frac 32}
\end{align}
with $\theta (x)$ the step function. The $M^2_{BR}$ value depends on the particle under consideration and its main decay channel. 
We must consider particles with vanishing isospin to compute the $\Pi^{88}_{V}$. For the lightest isoscalar resonances, one can use the standard decomposition
\begin{align}
\omega_8=\sqrt{\frac 13}\omega -\sqrt{\frac 23}\phi\qquad\rightarrow\qquad \Pi_V^{88}=\frac 13 \Pi_V^\omega+\frac 23 \Pi_V^\phi.
\end{align}
Similarly, the U-spin axial spectral function is
\begin{align}
   \Pi_A^{U}=\Pi_A^{K_1(1270)} +\Pi_A^{K_1(1400)} 
\end{align}
and involves the $\frac 12 (1^+)$ resonances  $K_1(1270)$ and $K_1(1400)$. Furthermore, we will consider all additional resonances with the appropriate quantum numbers listed by the Particle Data Group \cite{pdg} in the pertinent spectral functions.

For the lightest resonances, that is $\omega$ and $\phi$, the constants $f_\omega$ and $f_\phi$,  are fixed by  the combined use of the two
KFSR relations for the  $\rho$ meson~\cite{Kawarabayashi:1966kd,Riazuddin:1966sw},  yielding for the (Sakurai) $\rho$-photon mixing coupling
$g_{\gamma\rho}=-f_\rho m_\rho$. Assuming this to hold for the lowest $1^{--}$  vector
couplings $g_\omega, g_\phi$ as well, and recalling  that in  Hidden local symmetry~\cite{Bando:1985rf,Klingl:1996by}, 
\bea
\frac{g_{\gamma\rho}}{m_\rho^2}=\frac {3g_{\gamma\omega}}{m_{\omega^2}}=-\frac{3 g_{\gamma\phi}}{\sqrt 2 m_\phi^2}
\eea
we obtain for the vector decays, the  ratios
\bea
\frac {f_\rho}{m_\rho}=3\frac{f_\omega}{m_\omega}=-\frac 3{\sqrt 2}\frac {f_\phi}{m_\phi}
\eea 
We have used $f_\rho=\sqrt{2} f_\pi$ with $f_\pi=93$ MeV. Instead, we have used $f=\Gamma_0$ for all heavier resonances, with $\Gamma_0$ reported in ref. \cite{pdg}. The mass $M_{BR}$ has been chosen from the main strong decay channel, favoring channels involving pions when no leading channel could be identified due to lack of data. Table \ref{tab:particles} summarizes the characteristics of the resonances included in this work. 
\begin{table}
    \centering
    \begin{tabular}{c c c c c c}
    \hline
    \hline
     & $I(J^{PC})$ & $m$[MeV]   &  $f$[MeV] & $\Gamma_0$[MeV] & $M_{BR}$\\
    \hline
    \hline
    $\omega$ & $0(1^{--})$ & 782.66 & $44.26$ & 8.68 & $3m_\pi$\\
    $\phi$ & $0(1^{--})$ & 1019.461 & $-81.53$ & 4.249 & $2m_K$\\
    $\omega(1420)$ & $0(1^{--})$ & 1410 & $290$ & 290 & $m_\rho+m_\pi$\\
    $\phi(1680)$ & $0(1^{--})$ & 1680 & $150$ & 150 & $2m_K+m_\pi$\\
    $\omega(1650)$ & $0(1^{--})$ & 1670 & $315$ & 315 & $m_\pi+m_\rho$\\
    $\phi(2170)$ & $0(1^{--})$ & 2164 & $106$ & 106 & $m_\phi+2m_\pi$\\
    $\omega(2220)$ & $0(1^{--})$ & 2232 & $93$ & 93 & $2m_\pi+m_\omega$\\
    \hline
    $K_1$(1270) & $\frac{1}{2}(1^{+})$ &1253 & 90& 90& $m_\rho+m_K$ \\
    $K_1$(1400) & $\frac{1}{2}(1^{+})$ &1403 & 174& 174& $m_{K^*(892)}+m_\pi$ \\
    $K_1$(1670) & $\frac{1}{2}(1^{+})$ &1650 & 150& 150& $m_{K}+2m_\pi$ \\
    \hline
    \hline
    \end{tabular}
    \caption{The properties of particles appearing in the relevant spectral functions. The decay constants of $\omega$ and $\phi$ are obtained using the KFSR relations, whereas for the other resonances, we have used $f=\Gamma_0$. Masses and decay widths are taken from ref.~\cite{pdg}.}
    \label{tab:particles}
\end{table}
We also mention here that the adimensional coupling $G_V$ is given by $G_V=m_\phi^2/g_{\gamma\phi}$, even though this value is irrelevant for $\rho_{00}$, as it cancels in the ratio.

In Fig.~\ref{fig:spectralfunction}, we show the 88-vector (left) and 88-axial-vector (right) spectral functions, following from the parametrizations (\ref{PIVA}) with
the resonance parameters listed in table \ref{tab:particles}. The vector spectral function is dominant, which implies a smaller ``chiral mixing'' by thermal kaons in the $\phi$ channel. This is in contrast to the $\rho$-channel, where the ``chiral mixing'' by thermal pions is a major source of chiral restoration~\cite{Steele:1996su,Dusling:2006yv,Dusling:2007su}. 

Indeed, the mixing of the vector-axial spectral functions caused by the kaon rescattering can be estimated by retaining the forward contributions in (\ref{eq:W0 res gas}) and (\ref{eq:rate1}). Approximating the integrals in the kaon momentum as diluteness factors, one gets 
\bea
&& 2E_q(2\pi)^3\bigg(\frac{dR^0}{d^3 q }  
+\frac{dR^1_{L,\perp}}{d^3 q }\bigg)=\bigg(\frac{1}{e^{\beta E^\phi_q}+1}\frac{G_V^4f^2_\phi}{m_\phi^2} \bigg)\nonumber\\
&&\times -2q^2\epsilon^\mu\epsilon_\mu^*\bigg({\rm Im}\Pi_V^\phi(q)
-\,4\kappa_K\big({\rm Im}\Pi_V^{88}(q)-{\rm Im}\Pi_A^U(q)\big)\bigg).
\eea
The mixing in the spectral functions in the bracket is described by an \emph{in-medium spectral function}, which reads:
\bea
\label{PITX}
{\rm Im}\Pi_T(q) &= \big({\rm Im}\Pi_V^\phi(q)
-4\kappa_K{\rm Im}\Pi_V^{88}(q\big))+4\kappa_K{\rm Im}\Pi_A^U(q)\nonumber
\\
&\sim
\bigg(1-\frac 83\kappa_K\bigg){\rm Im}\Pi_V^\phi(q)+4\kappa_K{\rm Im}\Pi_A^U(q)
\eea
\begin{figure}
    \centering
    \includegraphics[scale=1]{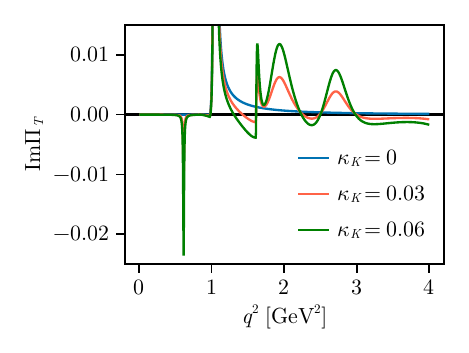}
    \caption{The effective in-medium spectral function for different values of $\kappa_K$.}
    \label{fig:thermal spectral function}
\end{figure}
The ``chiral mixing'' is total when the diluteness parameter reaches $4\kappa_K=\frac 35$, which is rather large. For instance, if $\mu=0$, then $\kappa_K\simeq 0.6$ is achieved for an unrealistic hadronic temperature of $T=650\; \mathrm{MeV}$. So, we expect the corrections to be small without a finite chemical potential. However, the correction will become larger if the kaons are assumed out of chemical equilibrium and a non-vanishing chemical potential is introduced.
In figure \ref{fig:thermal spectral function}, we show the in-medium spectral function (\ref{PITX}) for various values of the kaon diluteness parameter. For example, at a temperature of 170 MeV, the values of $\kappa_K\simeq 0.03$ and $\kappa_K\simeq 0.06$ can be achieved for $\mu/T=0.8$ and $\mu/T=1.3$, respectively.

\subsection{Static fluid in thermal equilibrium production}
\begin{figure}
    \centering
    \includegraphics{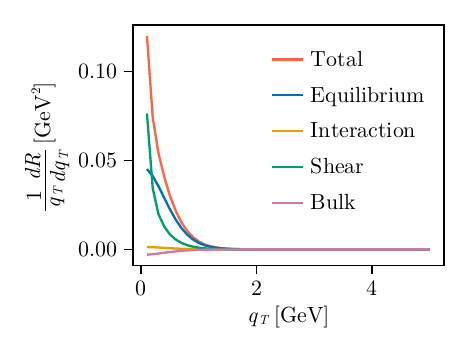}
    \includegraphics[width=0.49\textwidth]{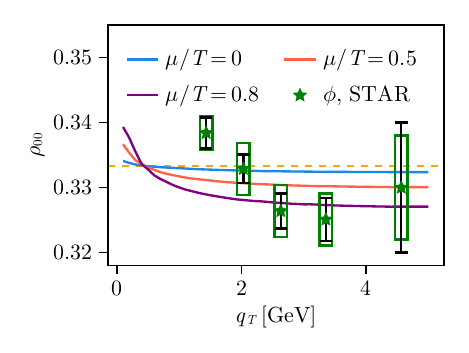}
    \caption{Left panel: the unnormalized emission rate of $\phi$ meson in a static fluid, including different contributions. The dissipative coefficients are set to $\zeta/s=\eta/s=1$ and $\mu/T=0$. Right panel: The $00$-element of the spin density matrix in a static background. The dissipative coefficients are $\zeta/s=\eta/s=0.1$. For $T=170$ MeV, the values of the chemical potentials $\mu/T=\{0, 0.5,0.8\}$ correspond to $\kappa_K\simeq\{0.01,0.02, 0.03\}$, respectively.}
    \label{fig:static}
\end{figure}
In this section, we compute the rates of a static fluid in thermal equilibrium with a constant temperature $T=T_0$ and four-velocity $u^\mu=(1,0,0,0)$ in Minkowski coordinates. Here we choose $T_0=170$ MeV. In the static limit, the dissipative corrections are not expected to contribute, as $\theta=\sigma_{\mu\nu}=0$. However, for illustrative purposes, we choose $\sigma_{\mu\nu}\hat{q}^\mu \hat{q}^\nu =1$ and $\theta=1$. This choice is made to understand the effect of dissipation qualitatively. A more sensible approach will be adopted in the next section.

The left panel of figure \ref{fig:static} shows the unnormalized production rate per unit of transverse momentum, together with the contributions from the ``equilibrium'', the interaction, the shear, and the bulk viscosity terms. The shear and bulk viscosity ratio to entropy density has been set to $\eta/s=\zeta/s=1$. These values are unrealistic but help us better visualize the contribution of dissipation, and once again, they are chosen for illustrative purposes. We also chose a chemical potential $\mu/T_0 = 0.8$, corresponding to $\kappa_K\simeq 0.03$.
The rate is dominated by the equilibrium and shear terms, with other corrections almost negligible. This is due to the very small diluteness of kaons and the spectral functions.

The right panel of figure \ref{fig:static} shows the $\rho_{00}$ element of the spin density matrix for different values of $\mu/T$, with a dashed line representing $\rho_{00}=1/3$, the isotropic value. For this plot, $T_0=170$ MeV, and we choose $\eta/s=\zeta/s=0.1$. We can see that the predicted $\rho_{00}$ is compatible with $1/3$, and with the experimental data from \cite{STAR:2022fan}, even in this oversimplified case.

\subsection{Off-equilibirum production: Bjorken flow}
\begin{figure}
    \centering
    \includegraphics[]{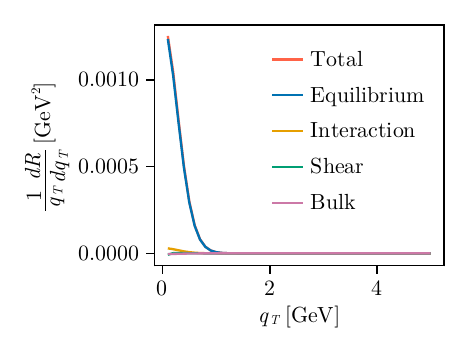}
    \includegraphics[width=0.49\textwidth]{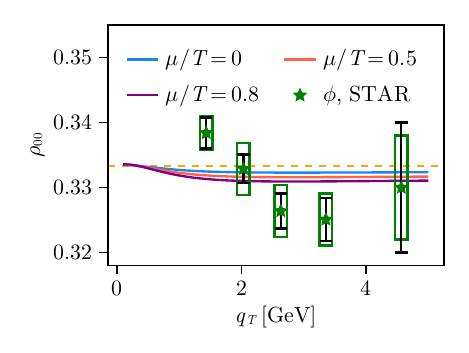}
    \caption{Left panel: the unnormalized emission rate of $\phi$ meson in a background Bjorken flow, including different contributions. Right panel: The $00$-element of the spin density matrix in a Bjorken flow for different values of $\mu/T$. The dissipative coefficients are $\zeta/s=\eta/s=0.1$ in the right panel, and to $1$ in the left panel.}
    \label{fig:bjk}
\end{figure}
Here we move to a more realistic case and compute the rates for a fluid obeying Bjorken flow. For this purpose, it is convenient to use Milne coordinates, defining the proper time and the rapidity as $\tau=\sqrt{t^2-z^2}$ and $\eta=1/2\ln[(t+z)/(t-z)]$. In this case, the temperature, the expansion scalar, and the shear tensor (in Milne coordinates) have the following expressions:
\begin{equation}
    T=T_0\left(\frac{\tau_0}{\tau}\right)^{1/3}, \qquad \theta=\frac{1}{\tau},\qquad 
    \sigma_{\mu\nu}={\rm diag}\left\{0,\frac{1}{3\tau},\frac{r^2}{3\tau},-\frac{2\tau}{3}\right\}.
\end{equation} 
The flow velocity in Milne coordinates reads $u^\mu=(1,0,0,0)$. In this case, since our model applies in the hadronic phase, we expect the result to be less dependent on the value of the free parameters  $\eta/s$ and $\zeta/s$, because the kinematic quantities $\theta$ and $\sigma_{\mu\nu}\hat{q}^\mu\hat{q}^\nu$ will drop with proper time. For our calculations, we have assumed that the hydrodynamic stage starts at $\tau_0^{(hyd.)}=0.5$ fm at a temperature of $T_0^{(hyd.)}=300$ MeV, whereas the hadronic phase is achieved when the temperature is $T^{(had.)}=170$ MeV. The freeze-out occurs at $T^{f.o.}=145$ MeV. These values determine the proper times to be used in the volume integrals associated with the rate. Thanks to the symmetries of the setup, the radial and azimuthal integrals factor out, and the integral in the spatial rapidity is taken in the interval $\eta\in(-1,1)$.

The left panel of figure \ref{fig:bjk} shows the unnormalized production rate per unit of transverse momentum, together with the single contributions from the equilibrium, the interaction, the shear, and the bulk viscosity terms. We have used $\zeta/s=\eta/s=1$ and $\mu/T=0.8$, but even for these values, we can see that the viscosity contribution to the emission rate is much smaller than the previous section, in agreement with our expectations.  

Finally, the right panel of figure \ref{fig:bjk} represents the $\rho_{00}$ element of the spin density matrix for different values of chemical potential, together with experimental data from \cite{STAR:2022fan}. For this plot $\zeta/s=\eta/s=0.1$, the result doesn't change significantly even for $\zeta/s=\eta/s=1$. We can see that varying the chemical potential has a smaller effect here because the contribution from viscosities is suppressed. As for the static background, even in this case, the calculations are consistent with $1/3$ and in qualitative agreement with the data. 

\section{Conclusions}
\label{SEC5}
We have analyzed the emission of polarized $\phi$ mesons in an interacting hadronic fluid, 
in-, and out-of-equilibrium, including the effects of viscosities and flow. Interactions with the thermal medium and non-equilibrium flow break the isotropy of the spin distribution, leading to an imbalance of longitudinally and transversely polarized $\phi$ mesons. These effects have been studied using the vector dominance model and the chiral reduction scheme developed in \cite{Yamagishi:1995kr}, where the corrections due to interactions and fluid viscosities and flow are expressed through pertinent spectral functions. We stress that the chiral constraints or Ward identities used in our analysis should be enforced in any analysis of the
hadronic emissivities at current collider energies.

The ``chiral mixing'' through thermal kaons in the $\phi$-channel is very small compared to that observed in the $\rho$-channel through thermal pions. Indeed, the $\phi$ rates are dominated by the equilibrium rates, with small corrections from the hadronic interactions to linear order in the thermal kaon densities. This is mostly due to the small kaon diluteness corrections in the typical range of hadronic temperatures probed by the RHIC collisions. The presence of a chemical potential can increase the diluteness, enhancing the effect.

 To compare the polarized $\phi$-meson emissivities to those reported at RHIC, we have modeled the fluid as a Bjorken flow. Our results for $\rho_{00}$ as a function of $p_T$ show small
 deviations from the isotropic baseline at intermediate momentum, whereas at small momenta the deviation wanes off. The effect is strongly suppressed if no chemical potential in introduced, see previous discussion about the diluteness of kaons. Our results are qualitatively consistent with the recently reported data by the STAR collaboration. 
 
 This analysis shows
 that the leading thermal $\phi$-emissivities are robust against higher-order interactions and viscous corrections. Extension to lower energy collisions, where the deviation from the isotropic polarization observed experimentally is more sizable, can be carried out considering additional corrections from baryon states. We will address this issue next.

{\bf Acknowledgements} A. Palermo is grateful to F. He for enlightening discussions. This work is supported by the Office of Science, U.S. Department of Energy under Contract No. DE-FG-88ER40388.

\appendix

\section{Evaluating the interaction and dissipation integrals}\label{app:integrals}
The numerical evaluation of \eqref{RATEE} is costly,  mainly due to the presence of the principal value. To simplify the problem, it is convenient to boost the integration variable to the rest frame of the $\phi$, i.e. the frame where the momentum of the $\phi$ is $\mathfrak{q}=(m_\phi,0,0,0)$. Denoting $\Lambda$ the boost such that $q=\Lambda \mathfrak{q}$ and writing $p=\Lambda k$ we have
\begin{align}
\frac{dR^1_{L,\perp}}{d^3 q }  =&\frac{1}{e^{\beta E^\phi_q}+1}\frac{g_V^4f^2_\phi}{m_\phi^2}  \frac{ 1  }{ (2\pi)^3 E^\phi_q } \,
\int \frac{d^3k}{(2\pi)^3}\frac 1{2E^K_k}\frac 1{e^{\beta'\cdot k}-1}\nonumber\\
&\times\bigg[\frac {2q^2}{f_K^2}\, {\rm Im}\Pi_V^{88}(q)
-((m_K^2+m_\phi^2\pm2E_k^K m_\phi)-|\epsilon^{RF}_{L,\perp}\cdot k|^2)\frac 3{4f_K^2}\,  {\rm Im}\Pi_A^{88}(m_K^2+m_\phi^2\pm 2E_k^Km_\phi)\nonumber\\
&-q^2|\epsilon^{RF}_{L,\perp}\cdot k|^2\frac{32}{f_K^2}\,{\rm Re}\,\Delta_R(m_K^2+m_\phi^2\pm 2E_k^Km_\phi)\, {\rm Im}\Pi_V^{88}(q)\bigg]_{q^2=m_\phi^2}
 \end{align}
 where $\beta'=\Lambda^{-1}\beta$, with $\beta^\mu=u^\mu/T$, and
 \begin{equation*}
     \Lambda^{\mu}_{\nu}=\frac{1}{m_\phi}
     \begin{pmatrix}
        q^0 & q^1 & q^2 & q^3\\
        q^1 & 1+\frac{{q^1}^2}{m+q^0} & \frac{q^1 q^2}{m+q^0} & \frac{q^1 q^3}{m+q^0}\\
        q^1 & \frac{q^1 q^2}{m+q^0} & 1+\frac{{q^2}^2}{m+q^0} & \frac{q^2 q^3}{m+q^0}\\
        q^1 & \frac{q^1 q^3}{m+q^0} & \frac{q^2 q^3}{m+q^0} & 1+\frac{{q^3}^2}{m+q^0}\\
     \end{pmatrix}.
 \end{equation*}
 In the rest frame, the polarization vectors read:
 \begin{equation}
     {\epsilon^{RF}}^\mu_{-1}=\frac{1}{\sqrt{2}}(0,1,-i,0),\qquad {\epsilon^{RF}}^\mu_{0}=(0,0,0,1),
     \qquad
     {\epsilon^{RF}}^\mu_{1}=-\frac{1}{\sqrt{2}}(0,1,i,0).
 \end{equation}
 Then, the principal value is rewritten
\begin{equation*}
 {\rm Re}\,\Delta_R(p\pm q)=\frac{1}{2p\cdot q+m_\phi^2}+\frac{\rm PP}{-2p\cdot q+m_\phi^2}=\frac{1}{2E_k^Km_\phi+m_\phi^2}+\frac{\rm PP}{-2E_k^Km_\phi+m_\phi^2}.
\end{equation*}

The denominator of the last term vanishes for
$$k=\frac{\sqrt{m_\phi^2-4m_K^2}}{2}$$
which is the cutting point for the evaluation of the principal value.  Thanks to the variable change, the principal part has been reduced to one variable integral. 

Moving to the dissipative corrections, eq. \eqref{RATEX}, the integral is written in covariant form as:
 \begin{align}
     \int \frac{d^3p}{(2\pi)^3}\frac 1{2E^K_p}\frac {e^{\beta p\cdot u}}{(e^{\beta p\cdot u}-1)^2}\frac{p_\mu p_\nu\Delta^{\mu\rho}\Delta^{\nu\sigma}\partial_\rho\beta_\sigma}{p\cdot u}f(p,q,u),
 \end{align}
 where $f(p,q,u)$ is the term in square brackets.
 Defining $\nabla_\mu = \Delta_{\mu\nu} \partial^\nu$, with $\Delta^{\mu\nu}=g^{\mu\nu}-u^\mu u^\nu$, and recalling that $u^\mu=T\beta^\mu$, one has:
 \begin{equation}
     \Delta_{\mu\nu}\nabla_\rho \beta^\nu = \frac{1}{T}\nabla_\rho u_\mu
 \end{equation}
 the numerator can be decomposed into shear and bulk parts
  \begin{align}
    \frac{1}{T} \int \frac{d^3p}{(2\pi)^3}\frac 1{2E^K_p}\frac {e^{\beta p\cdot u}}{(e^{\beta p\cdot u}-1)^2}\frac{p_\mu p_\nu(\sigma^{\mu\nu}+\frac{1}{3}\theta\Delta^{\mu\nu})}{p\cdot u}f(p,q,u)
 \end{align}
 with
 \begin{equation*}
     \sigma^{\mu\nu}=\frac{1}{2}\left(\nabla^\mu u^\nu+\nabla^\nu u^\mu\right)-\frac{1}{3}\theta\Delta^{\mu\nu}, \qquad \qquad \theta=\nabla^\mu u_\mu.
 \end{equation*}
 To simplify the integration further, we decompose the integral as
\begin{equation}
    \int \di^3 p \,p^\mu p^\nu F(p,q,u)=A g^{\mu\nu}+Bq^\mu q^\nu +C(u^\mu q^\nu +u^\nu q^\mu)+D u^\mu u^\nu.
\end{equation}
Notice that $C$ and $D$ will not contribute to the contraction with $\sigma_{\mu\nu}$ and $\Delta_{\mu\nu}$.
Contracting the above equation with all the possible combinations of $g^{\mu\nu}$, $q^\mu$ and $u^\mu$, we define $x$, $y$, $z$ and $w$ and obtain 
\begin{align*}
    x&\equiv m_K^2\int\di^3p F=4A+m_\phi^2 B +2u\cdot q C +D\\
    y&\equiv \int\di^3p(p\cdot q)^2 F= A m_\phi^2+Bm_\phi^4+2u\cdot q m_\phi^2+D(u\cdot q)^2\\
    z&\equiv \int\di^3p (p\cdot u)^2 F=A+B (u\cdot q)^2+2C(u\cdot q)+D\\
    w&\equiv \int \di^3p (p\cdot u)(p\cdot q) F=A u\cdot q+Bm_\phi^2(u\cdot q)+(m_\phi^2+(u\cdot q)^2)+D(u\cdot q)\\
\end{align*}
The solutions to the system of equations are
\begin{align*}
    A&=-\frac{q^2 (-x)+q^2 z+(u\cdot q)^2 x-2
   u\cdot q w+y}{2 \left(q^2-(u\cdot q)^2\right)}\\
   B&=-\frac{q^2
   x-q^2 z-(u\cdot q)^2 x-2 (u\cdot q)^2 z+6 u\cdot q w-3 y}{2
   \left(q^2-(u\cdot q)^2\right)^2}\\
   C&=-\frac{-q^2 u\cdot q x+3
   q^2 u\cdot q z-2 q^2 w-4 (u\cdot q)^2 w+u\cdot q^3 x+3
   u\cdot q y}{2 \left(q^2-(u\cdot q)^2\right)^2}\\
   D&=  -\frac{-q^2 (u\cdot q)^2 x+6 q^2 u\cdot q w+q^4 x-q^2 y-3 q^4
   z-2 (u\cdot q)^2 y}{2
   \left(q^2-(u\cdot q)^2\right)^2}
\end{align*}
with finally
\begin{equation}
    \int \di^3p\, p^\mu p^\nu F(p,q,u) (\sigma_{\mu\nu}+\frac{1}{3}\theta\Delta_{\mu\nu})=Bq_{\mu}q_{\nu}\sigma^{\mu\nu}+\theta(A+\frac{B}{3}(q^2-(u\cdot q)^2))
\end{equation}
The same technique involving the boost to the $\phi$ rest frame as presented earlier can now be used to compute $x,y,z,w$, and $A$ and $B$ thereof. 

\bibliography{biblio}

\end{document}